# Entanglement-based quantum deep learning

**Zhenwei Yang and Xiangdong Zhang**

Key Laboratory of advanced optoelectronic quantum architecture and measurements of Ministry of Education, School of Physics, Beijing Institute of Technology, 100081, Beijing, China

**Abstract**

Classical deep learning algorithms have aroused great interest in both academia and industry for their utility in image recognition, language translation, decision-making problems and more. In this work, we have provided a quantum deep learning scheme based on multi-qubit entanglement states, including computation and training of neural network in full quantum process. In the course of training, efficient calculation of the distance between unknown unit vector and known unit vector has been realized by proper measurement based on the Greenberger-Horne-Zeilinger entanglement states. An exponential speedup over classical algorithms has been demonstrated. In the process of computation, quantum scheme corresponding to multi-layer feedforward neural network has been provided. We have shown the utility of our scheme using Iris dataset. The extensibility of the present scheme to different types of model has also been analyzed

## 1. Introduction

Machine learning, as an interdisciplinary subject in the fields of computer science, mathematics, statistics and neuroscience, has made outstanding achievements in recent years. Using such an approach, many domains such as speech recognition, visual object recognition, object detection have been innovated [1-2]. Deep learning based on neural network is one branch of machine learning and has attracted much attention as it can solve complicated practical tasks very well [3-5]. It is trained by a lot of data and aims to obtain optimal parameters of neural network. In order to solve various tasks, many models of neural network have been constructed [6-13], such as multi-layer feedforward [6], convolutional [7], Elman [8] and Hopfield neural networks [9]. Multi-layer feedforward and convolutional neural networks can be applied to the task of classifying and pattern recognition [14-17]; Elman and Hopfield neural networks can be applied to the task of reconstruction as a memory system [9,18]. It is noteworthy that multi-layer feedforward neural network (MFNN) is a basic model. Studying this model not only promotes our understanding on the mechanism of neural network, but also inspires many other models to be constructed.

On the other hand, quantum computation is appreciated in recent decades as it can employ uniquely characteristic of entanglement to speed up algorithms for many tasks. Related examples have been demonstrated in many experiments of various quantum systems [19-22]. If quantum computation can be combined with machine learning, typically with deep learning, to construct corresponding quantum learning algorithms, it is no doubt that they should have both advantages of machine learning and quantum computation. Thus, such a topic has attracted a lot of research interest. Now, some quantum machine learning algorithms have been proposed and demonstrated [23-33], such as quantum support vector machine [25], supervised and unsupervised machine learning [23], quantum-enhanced machine learning [27], and distributed quantum learning [33]. Based on these algorithms, some schemes of quantum deep learning (QDL) have been discussed [34-44]. Quantum Boltzmann machine provides efficient training and a more comprehensive framework than classical counterpart [34-36]. At the same time, D-wave quantum annealer has been proposed as a potential physical implementation for Quantum Boltzmann machine and some other quantum-assisted machine learning algorithms [37-38]. Quantum Hopfield neural network based on the simulation of sparse Hamiltonian and the solution of linear systems of equations has been discussed to realize exponential speed up in computation and training [39]. Quantum convolutional neural network based on the reverse-direction correspondence of the multi-scale entanglement renormalization ansatz has been analyzed to realize efficient computation by exponentially reducing the number of parameters [42]. However, these schemes only focus on special models for particular tasks. The question is whether a more general and efficient scheme of QDL can be constructed to solve more tasks in full quantum process. Here, the full quantum process means that both computation and training based on neural networks are quantum processes.

In fact, we notice that three general schemes based on MFNN have been proposed. One scheme encodes data into relative phases of wavefunctions and realizes the Baqprop principle, which corresponds to the error Back Propagation algorithm in classical neural network [43]. Another scheme focuses on a quantum optical neural network that encodes data into Fock states [44]. However, these two ways of encoding are not often used in conventional quantum computations. A conventional way is that encoding data into the probability amplitudes of a quantum state, which is in the Hilbert space constructed by the way of tensor product of qubits. Thus, the above two schemes may need special conversion procedures to cooperate with some current quantum setups and algorithms. The third scheme is a quantum generalization of MFNN [45]. Although the conventional encoding way is used, the required qubits are polynomially increased with the number of neurons as one ancillary qubit is used to complete the function of one neuron.

In this work, we propose a new general scheme of QDL that encodes data into the probability amplitudes of quantum state. In our scheme, the form of vector is used and full quantum process is realized using quantum MFNN (QMFNN). The required qubits in our scheme increase logarithmically with the number of neurons. The scheme includes two parts: quantum neural network (QNN) and training. In the part of QNN, our scheme is applicable to different types of model such as fully connected network, quantum convolutional neural network and so on, and here the QMFNN is used to demonstrate the function as an example. In the part of training, we realize exponential speed up for the computation of loss function by taking advantage of entanglement. The speed up is from the measurements corresponding to the distance between a known unit vector and an unknown unit vector. Using Iris dataset, we demonstrate the feasibility of our scheme in the way of numerical simulation. A large proportion of samples in dataset are successfully recognized. Finally, we discuss the efficiency of QMFNN and training. Besides, the compatibility of this scheme is also discussed that other models of QNN can be combined with our scheme to realize full quantum process and more efficient computation.

## 2. Scheme of quantum deep learning

The MFNN is a basic model of classical deep learning as shown in Fig. 1(A). It includes an input layer, some hidden layers and an output layer, which can be represented by column vector $\vec{X}$, $\vec{Y}$ and $\vec{Z}$, respectively. Every layer is connected by weight matrix $W$. If $N$ neurons are included in every layer, the data of input layer is described by $\vec{X}=(x_1 \ x_2 \ ... \ x_N)^T$. Other vectors are written similarly. The $W^1$ is the weight matrix between input layer and the first hidden layer, matrix element $w_{ij}^1$ represents the weight connection between the $j_{th}$ neuron of input layer and the $i_{th}$ neuron of the first hidden layer. The information received by the first hidden layer is $\vec{H} = W^1 \vec{X}$. After the nonlinear transformation in the hidden layer, the output is expressed as $\vec{Y}$. Here $y_i = f(h_i)$, $f$ is the nonlinear function, $i$ $(=1, 2, ..., N)$ represents the $i_{th}$ neuron of the hidden layer. The same processes appear when the information passes through other hidden layers, weight matrices and output layer. Finally, the output $\vec{Z}$ is obtained. Besides the above computation of neural network, training (or learning) is important for deep learning. Using the output $\vec{Z}$ and the label $\vec{S}$ of the sample, loss function is calculated to characterize the difference between $\vec{Z}$ and $\vec{S}$. A small difference is usually required. Mean square error (MSE) $E = \sum_{i=1}^{N}(z_i - s_i)^2$ is frequently used as loss function, $s_i$ represents the $i_{th}$ element of $\vec{S}$. Combined with the learning algorithm, like backpropagating (BP) error algorithm [46], optimized weight matrices can be obtained and the training is completed. The neural network after training can work well on other similar data.

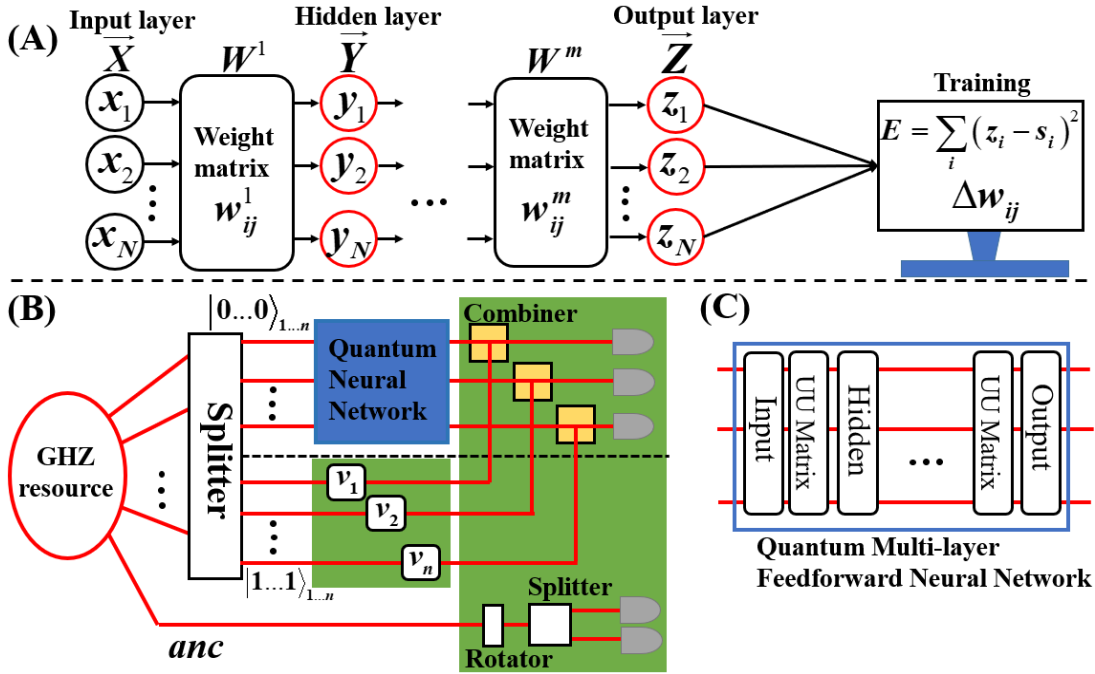

**Figure** 1. (A) Classical multi-layer feedforward neural network, including input layer, some hidden layers and output layer from left to right. Every layer is connected by weight matrix. Generally training is completed by computer, which includes the computation of mean square error $E$ and the gradient $\Delta w_{ij}$ of the parameters of weight matrix. (B) The scheme of quantum deep learning. Starting from the Greenberger-Horne-Zeilinger entanglement resource, ancillary qubit represented by anc is measured in some bases. Other qubits are processed by quantum neural network in blue box and sample labels in the left green box respectively according to $|0...0\rangle_{1...n}$ and $|1...1\rangle_{1...n}$. $v_i (i=1,...,n)$ is single-qubit unitary transformation for the $i_{th}$ qubit. Quantum multi-layer feedforward neural network as an example of quantum neural network is constructed by input layer, hidden layers and output layer shown in Fig. 1(C), Input, UU Matrix and Output represent input layer, universal unitary matrix and output layer corresponding to the structure of Fig. 1(A). Measurements are realized in the right green box. Measurements of different bases for ancillary qubit are achieved by rotator. The splitter makes qubit be separated into different paths according to the bases $|0\rangle$ and $|1\rangle$. The combiner makes qubit superposed in different paths be injected into same detector. The training is completed by the two green boxes.

Corresponding to the classical deep learning scheme based on the MFNN, we propose a general scheme of QDL consisting of QNN and training as shown in Fig. 1(B). Blue shaded box in Fig. 1(B) represents the QNN, and the details of QMFNN are shown in Fig. 1(C) as an example, which includes input layer, universal unitary (UU) matrix, hidden layer and output layer corresponding to the structure of neural network in Fig. 1(A). Two green shaded boxes are the part of training, which include sample labels in the left box and measurements in the right box corresponding to the function of computer in Fig. 1(A). Our scheme is based on the Greenberger-Horne-Zeilinger (GHZ) entanglement resource. The GHZ state with $n+1$ qubits can be written as

$$|\psi\rangle = \frac{1}{\sqrt{2}}\left(|0\rangle_{anc}|0...0\rangle_{1...n} + |1\rangle_{anc}|1...1\rangle_{1...n}\right). \tag{1}$$

Here, ancillary qubit is represented by anc and directed into measurement. Other qubits are represented by number 1, 2…, n and split into different paths according to the sub-state $|0...0\rangle_{1...n}$ and $|1...1\rangle_{1...n}$ when they pass through the splitter. In real physical systems, such as quantum optics system with polarization, the function of splitter can be realized by polarized beam splitter (PBS). In Fig. 1(B), the upper half of dotted black line represents the paths for $|0...0\rangle_{1...n}$ and the lower part for $|1...1\rangle_{1...n}$. Then, the sub-state $|0...0\rangle_{1...n}$ enters the QNN and $|1...1\rangle_{1...n}$ enters the sample labels. In the following, we discuss the transformation of QMFNN in detail and how to implement training.

In the part of QMFNN as shown in Fig. 1(C), the sub-state $|0...0\rangle_{1...n}$ is transformed by the input layer, some UU matrices and output layer from left to right. First of all, we encode the sample data into quantum states in the input layer. Quantum states can be written as $|x\rangle_{1...n} = \frac{1}{A}\sum_{i=1}^{N} x_i |i\rangle$, here $|i\rangle$ is computational basis, $x_i/A$ is the probability amplitude of related basis and $A = \sqrt{\sum_{i=1}^{N}|x_i|^2}$ is normalized factor. So the sample data can be represented by probability amplitude. For the sample data $\vec{X}$ of N dimensions, we can efficiently prepare a sub-state $|x\rangle_{1...n}$ from $|0...0\rangle_{1...n}$ using qubits of $n = \log_2 N$. The techniques of quantum random access memory [47] and oracle-based quantum state preparation [48] can be used in this process. Besides the above two techniques required auxiliary qubit, there is also one method that does not require auxiliary qubit [49]. As we know, qubits are important resource in quantum computation and in such a case, the qubit overhead scales logarithmically with the dimension of data. This shows a logarithmical economy in the number of qubits compared with the scheme of [45], which uses one extra qubit to realize the function of one neuron. Then, the sub-state $|x\rangle_{1...n}$ is transformed by the UU matrix that is the weight matrix in the QMFNN and used to perform arbitrary unitary transformation. The UU matrix transforms $|x\rangle_{1...n}$ into $|u\rangle_{1...n}$. Now, the GHZ state is expressed as

$$|\psi\rangle = \frac{1}{\sqrt{2}}\left(|0\rangle_{anc}|u\rangle_{1...n} + |1\rangle_{anc}|1...1\rangle_{1...n}\right), \quad (2)$$

where $|u\rangle_{1...n} = \sum_{i=1}^{N}\alpha_i|i\rangle_{1...n}$, $\alpha_i$ is the probability amplitude of computational basis $|i\rangle_{1...n}$ and satisfies the condition $\sum_{i=1}^{N}|\alpha_i|^2 = 1$. Next, the sub-state $|u\rangle_{1...n}$ is nonlinearly transformed by the first hidden layer. Similar to the classical counterpart, the function of the hidden layer in the QMFNN is to realize one kind of nonlinear transformation. Thus, the probability amplitude $\alpha_i$ is changed into $\beta_i$. Assuming that $\sum_{i=1}^{N}|\beta_i|^2 = a$, $a > 0$ and $a$ is real. After normalization, we obtain the state

$$|\psi\rangle = \chi|0\rangle_{anc}|u'\rangle_{1...n} + \delta|1\rangle_{anc}|1...1\rangle_{1...n}, \quad (3)$$

$$|u'\rangle_{1...n} = \left(\beta_0|0\rangle_{1...n} + \beta_1|1\rangle_{1...n} + ... + \beta_{N-1}|N-1\rangle_{1...n}\right)/\sqrt{a}, \quad (4)$$

where $\chi = \sqrt{a}/\sqrt{a+1}$ and $\delta = 1/\sqrt{a+1}$. Similar transformation can be performed successively by using other UU matrices and hidden layers, and the function of output layer is similar to that of the hidden layer. After the transformation of QMFNN, the state of Eq. (3) is changed into

$$|\psi\rangle = \gamma|0\rangle_{anc}|k\rangle_{1...n} + \lambda|1\rangle_{anc}|1...1\rangle_{1...n}, \quad (5)$$

where $|k\rangle_{1...n}$ is the normalized output state of QMFNN. Through the above transformation process corresponding to the classical neural network, we can complete the computation of QMFNN.

In the part of training, the MSE and finite difference method are used as loss function and learning algorithm. In the following, we introduce how to realize these in the scheme. For the computation of MSE, the output of QMFNN and sample labels are required. In the part of QMFNN, the output is prepared as shown in Eq. (5). Sample labels can be completed by the left green box in Fig. 1(B), which contains some single-qubit unitary transformations $v_i$. When the sub-state $|1...1\rangle_{1...n}$ passes through the $v_i$, labels for different samples can be prepared and the state of Eq. (5) is changed into

$$|\psi\rangle = \gamma|0\rangle_{anc}|k\rangle_{1...n} + \lambda|1\rangle_{anc}|v\rangle_{1...n}, \quad (6)$$

where $|v\rangle_{1...n}$ represents the label of one sample, for example $|v\rangle_{12} = (1\ 0\ 0\ 0)^T$. In order to obtain MSE, two measurements are performed and the measurement setup is shown in the right green box of Fig. 1(B). The difference between two measurements lies in the choice of basis for the measurement of ancillary qubit. The choice can be realized by rotating rotator. Orthogonal measurement is realized when the ancillary qubit passes through the splitter and then is detected after the rotator. In the quantum optics system with polarization, the rotator and detector correspond to half-wave plate (HWP) and single-photon detector, respectively. As for other qubits, they are recombined on the combiner that connects two corresponding paths and then are detected. The combiner can be realized by beam splitter (BS) in reality.

First, we perform measurement on the bases of $|0\rangle$ and $|1\rangle$ for the ancillary qubit. The probabilities $\gamma^2$ and $\lambda^2$ in Eq. (6) can be obtained by measurement results of $|0\rangle$ and $|1\rangle$, respectively. According to the analyses of Eq. (3) and Eq. (4), $\gamma$ and $\lambda$ are real and thus can be obtained by the square root of $\gamma^2$ and $\lambda^2$. For the second measurement, we choose $|\varphi\rangle = \lambda|0\rangle - \gamma|1\rangle$ and $|\varphi_\perp\rangle = \gamma|0\rangle + \lambda|1\rangle$ as measurement bases for the ancillary qubit. Then Eq. (6) can be written in the bases of $|\varphi\rangle_{anc}$ and $|\varphi_\perp\rangle_{anc}$ as

$$|\psi\rangle = \gamma\lambda|\varphi\rangle_{anc}(|k\rangle - |v\rangle)_{1...n} + |\varphi_\perp\rangle_{anc}(\gamma^2|k\rangle + \lambda^2|v\rangle)_{1...n}. \quad (7)$$

And the probability that the ancillary qubit projects into $|\varphi\rangle$ is

$$P = \gamma^2 \lambda^2 E, \tag{8}$$

$$E = (|k\rangle - |v\rangle)^T (|k\rangle - |v\rangle), \tag{9}$$

where $E$ is the MSE between the output $|k\rangle$ of QMFNN and the label $|v\rangle$ of sample. And the MSE is equal to the square of the distance between two vectors. Using the probabilities $\gamma^2$ and $\lambda^2$ of the first measurement and the probability $P$ of the second measurement, the MSE is calculated by

$$E = \frac{P}{\gamma^2 \lambda^2}. \tag{10}$$

The accuracy of $E$ can be improved by increasing the times of repeated measurement. Here, the computation of MSE provides exponential speed up and the detailed discuss is given in Sec.4. The accumulated MSE $AccEk = \frac{1}{t}\sum_{i=1}^{t} E_i$ is used as final loss function for the dataset with $t$ samples.

In addition to the accumulated MSE obtained by the feedforward computation, deep learning also requires continuous updating of network parameters by learning algorithm. The BP algorithm is usually used in classical deep learning, but not for the QDL. The reason is that the computation of BP algorithm requires the output results of hidden layer, which is opposite to the quantum computation. In quantum computation, the result of middle process is not measured in general because the measurement can destroy the characteristic of entanglement. Considering that the gradients of the independent parameters of UU matrix are truly required in the training, we use finite difference method. Assuming that $\varsigma$ is an independent parameter of UU matrix and $\varepsilon$ is a tiny variable, the gradient can be calculated by

$$\Delta\varsigma = (AccEk(\varsigma + \varepsilon) - AccEk(\varsigma))/\varepsilon. \tag{11}$$

The gradient can be obtained by only two feedforward computations of accumulated MSE. In the process, quantum speed up is maintained. Then, the accumulated MSE decreases toward to the negative direction of gradient. Each parameter is updated as $\varsigma \leftarrow \varsigma - k\Delta\varsigma$, where $k$ is the learning rate.

Based on the measurement and calculation, the MSE and gradients can be efficiently obtained. After enough iterations, we can obtain a pretty small MSE and optimized UU matrices, which means that the training for QMFNN has been finished. Combined with the computation of QMFNN, full quantum process for deep learning has been implemented by our scheme. Such a scheme can be realized in various physical systems, such as quantum optics system. The GHZ resource can be prepared in the way of multi-photon or multi-degree of freedom. The UU matrix can be constructed by one-qubit and two-qubit gates using optical waveguide and linear elements. The measurement can be implemented by single photon detector with coincidence circuit. The above three parts have been demonstrated in many works [21, 50-53]. In fact, how to realize the nonlinear transformation is important in QMFNN. In the previous investigations, the ancillary qubit has been used to realize nonlinear interaction, like Knill-Laflamme-Milburn scheme [54]. However, these works aim to realize the unitary gate that is for the linear transformation. Besides, nonlinear quantum mechanics has been considered theoretically to imply polynomial-time solution for hard problems [55], it is very difficult to be realized experimentally. Here, we provide a way of realizing nonlinear transformation using quantum clone [56] and two-photon gate [57]. And the transformation is used to the experimental scheme of QDL, which is based on the degrees of freedom of polarization and orbital angular momentum (OAM) corresponding to three layers QMFNN with $N=4$. The detailed description of the QDL scheme including nonlinear transformation is shown in Appendix.

## 3. Numerical results of training and test

Deep learning is usually used to perform the classification that utilizes pattern recognition. In the field of quantum pattern cognition [58], Grover's search algorithm is used [59-60], which is a fixedly well-designed algorithm. In the following, we take the task of classification of Iris as an example to demonstrate the feasibility of our scheme in the way of numerical simulation. Here, feasibility means the ability of learning and generalization. Iris dataset as a frequently-used classifying experimental dataset contains three types (Setosa, Versicolour and Virginica) of 50 samples each. Each sample contains four attributes: Sepal Length (SL), Sepal Width (SW), Petal Length (PL) and Petal Width (PW). The dataset is usually divided into training set and testing set. The training set is constructed by randomly selecting 40 samples from each type and testing set is constructed by the others. We use training set to train 3-layer and 4-layer QMFNNs three times respectively, and then test the performance using testing set.

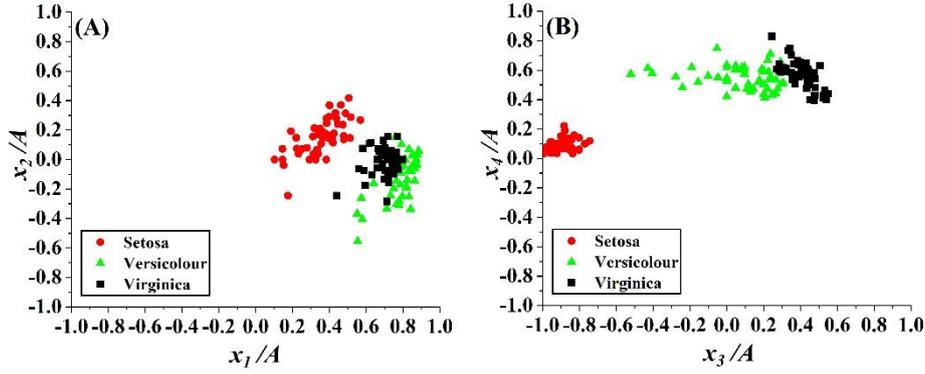

**Figure** 2. The results of pre-processing for initial data. The first two-dimensional data ($x_1/A$, $x_2/A$) and the latter two-dimensional data ($x_3/A$, $x_4/A$) of input state $|x\rangle_{in}$ are shown in (A) and (B) respectively. After normalization, the characteristic of Iris dataset is not changed. Versicolour's data are nonlinearly separated from Virginica's data. Setosa's data are linearly separated from the other two types. Red dots, green triangles and black boxes correspond to Setosa, Versicolour and Virginica, respectively.

In the input layer of QMFNN, attribute data of each sample are encoded into quantum state. We use $x_1$, $x_2$, $x_3$ and $x_4$ to represent SL, SW, PL and PW, respectively. According to the above method of normalization, the quantum sub-state for representing the sample data is $|x\rangle_{in} = (x_1 \ x_2 \ x_3 \ x_4)^T / A$, where $A = \sqrt{\sum_{i=1}^{4} x_i^2}$. The sub-state can be prepared by using unitary matrix to act on $|00\rangle$. If the data of Iris are simply normalized in the above manner, the states of some samples are highly similar and thus high accuracy of physical system is required. The reason is that the values for SL, SW and PL are too large to compare with the values of PL. In order to relax the requirement for future realization, a cut operation before normalization need be completed. The values for SL, SW and PL are cut by 4cm, 3cm and 4cm, respectively. In contrast, the values for PW are not changed. The results of pre-processing containing cut and normalization are shown in Fig. 2.

The first two-dimensional data ($x_1/A$, $x_2/A$) and the latter two-dimensional data ($x_3/A$, $x_4/A$) of input state $|x\rangle_{in}$ are shown in Fig. 2(A) and Fig. 2(B), respectively. Red dots correspond to Setosa, green triangles to Versicolour and black boxes to Virginica. It is seen that green triangles are cross with black boxes. This means that Versicolour's data are nonlinearly separated from Virginica's data.

Red dots are separated from the other two symbols. This means that Setosa's data are linearly separated from the other two types. Learning the ability of linear separation is not important in QDL, which can be realized by other algorithms such as quantum support vector machine [25]. Demonstration of the ability of nonlinear separation is the truly goal of the following training and test using our scheme. Here, the results of pre-processing show the nonlinear characteristic used.

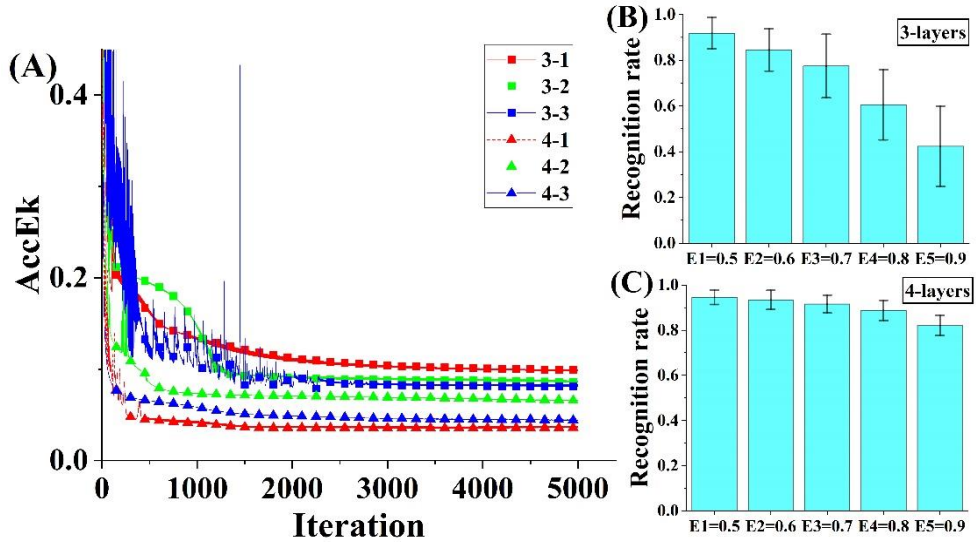

**Figure** 3. The results of training and test. (A) Accumulated MSE changes with the number of iterations. The marks 3-1, 3-2 and 3-3 correspond to three times of training for the case with 3-layer, respectively. The corresponding results for the case with 4-layer are marked by 4-1, 4-2 and 4-3. We do three times training for QMFNN with 3-layer and 4-layer. Each time contains 5000 iterations. Curves with boxes and triangles represent 3-layer and 4-layer cases, respectively. (B) and (C) show the recognition rate of 3-layer and 4-layer cases acting on testing set. E1, E2, E3, E4 and E5 are various thresholds of recognition, which is equal to 0.5, 0.6, 0.7, 0.8 and 0.9, respectively.

Now, we perform transformations of UU matrix and hidden layer. Here, the 4-dimensional real UU matrix is used. According to the method in Ref. [61], the 4-dimensional matrix can be constructed by 6 independent parameters in the way of meshing, the unit of parameter is taken as radian. At the beginning of training, these parameters are randomly assigned a value ranged from 0 to $2\pi$. Note that the complex unitary matrix can be used in principle. However, the mathematics of deep learning is usually real so we use real UU matrix in this demonstration. Nonlinear transformation is followed to perform in hidden layer. The transformation function is shown in Eq. (A1) and can be constructed using quantum clone setup and two-qubit gate. The detailed descriptions about the transformation function and construction are given in Appendix. Other UU matrices and nonlinear transformations are performed similarly.

Finally, after the nonlinear transformation is performed in the output layer, we can obtain the output sub-state $|k\rangle$ of QMFNN and thus the computation of QNN is completed. In summary, the transformations successively performed in QMFNNs with 3-layer and 4-layer are described by $Input \to UU \to NL \to UU \to NL$ and $Input \to UU \to NL \to UU \to NL \to UU \to NL$, respectively. The input represents the pre-processing in the input layer, UU is for the UU matrix transformation and NL represents the nonlinear transformation in hidden and output layers.

In the part of training, we make sub-state $|v\rangle$ in Eq. (6) be equal to $|v\rangle_1 = (1\ 0\ 0\ 0)^T$, $|v\rangle_2 = (0\ 1\ 0\ 0)^T$ and $|v\rangle_3 = (0\ 0\ 1\ 0)^T$ for labeling Setosa, Versicolour and Virginica, respectively. By the measurements descripted below Eq. (6), the accumulated MSE $AccEk$ can be obtained between the output of QMFNN and the sample label. Then, in order to update independent parameters of UU matrices, we change parameters with a tiny value $\varepsilon$ and recompute $AccEk$. According to Eq. (11), we can obtain the gradients $\Delta\varsigma$ of parameters and update parameters $\varsigma \leftarrow \varsigma - k\Delta\varsigma$. Here, $\varepsilon$ and $k$ are taken as 0.001rad and 0.05rad, respectively.

Using above settings, we do three times of training, and test the QMFNN with 3-layer and 4-layer, respectively. Each time of training contains 5000 iterations. The results of training are shown in Fig. 3(A). Curves with boxes and triangles represent the QMFNN with 3-layer and 4-layer, respectively. The marks 3-1, 3-2 and 3-3 correspond to three times of training for the case with 3-layer, respectively. The corresponding results for the case with 4-layer are marked by 4-1, 4-2 and 4-3. The $AccEk$ is decreased in general to a small value with the number of iterations. And the curves for 4-layer case are under the counterpart of 3-layer. Besides, the rates of convergence of $AccEk$ for 4-layer case are quicker than those for 3-layer. This means that firstly both trainings for 3-layer and 4-layer cases are well; secondly the training effect for the 4-layer case is better than that of 3-layer. This is because one more hidden layer brings more numbers of parameter and thus better ability of learning. Some techniques like the bold driver technique [62] can be used to obtain smaller MSE in the future.

After training, we use samples of testing set to test the performance of QMFNN. The performance is evaluated by comparing the fidelity between output state $|k\rangle$ and label state $|v\rangle_i \, (i=1,2,3)$ with the recognition threshold. The output state $|k\rangle$ is calculated by the optimal UU matrices, which are chosen corresponding to the minimum $AccEk$ in the training. The fidelity is calculated by $F = {}_i\langle v|k\rangle\langle k|v\rangle_i$. Recognition is successful when the fidelity is larger than the threshold. The test results for the QMFNN with 3-layer and 4-layer are shown in Fig. 3(B) and Fig. 3(C), respectively. The values for y-coordinate represent the recognition rate, which is defined as the radio of the number of successfully recognized samples and total test samples. The marks E1=0.5, E2=0.6, E3=0.7, E4=0.8 and E5=0.9 for x-coordinate correspond to different recognition thresholds. High recognition threshold is used in some rigorous situations, where high quality (similarity) of recognition is emphasized. In Fig. 3(B), the recognition rate of 3-layer QMFNN is fast decreased as the threshold increases. In contrast, the recognition rate is roughly stable for the 4-layer case as shown in Fig. 3(C). It proves that the network with a smaller MSE has well performance in applications.

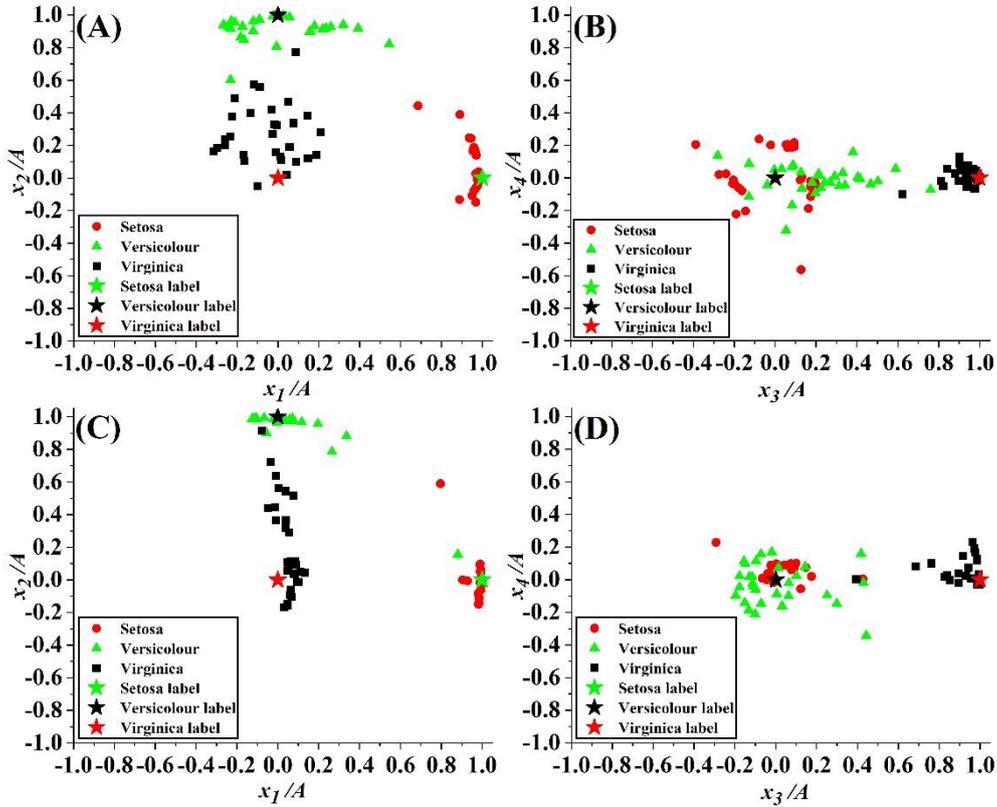

**Figure** 4. The test results of QMFNN with 3-layer case are shown in Fig. 4(A) and (B). The corresponding results for the 4-layer case are shown in Fig. 4(C) and (D). The first two dimensional data of results are shown in Fig. 4(A) and (C). The latter two dimensional data of results are shown in Fig. 4(B) and (D). Red dots, green triangles and black boxes correspond to Setosa, Versicolour and Virginica, respectively. Blue

star, black star and red star represent label states $|u\rangle_1$, $|u\rangle_2$ and $|u\rangle_3$ corresponding to Setosa, Versicolour and Virginica, respectively.

In Fig. 4, we show the output sub-state $|k\rangle$ of three times of test for 3-layer and 4-layer cases in detail. The test results for the 3-layer case are shown in Fig. 4(A) and (B), and the corresponding results for the 4-layer case are given in Fig. 4(C) and (D). Corresponding to Fig.2, the first two-dimensional data of results are shown in Fig. 4(A) and (C); and the latter two-dimensional data of results are shown in Fig. 4(B) and (D). The red dots, green triangles and black boxes also represent Setosa, Versicolour and Virginica, respectively. Besides, the blue star, black star and red star represent the labels for Setosa, Versicolour and Virginica, respectively, corresponding to the label states $|u\rangle_1$, $|u\rangle_2$ and $|u\rangle_3$. When the data of the output sub-states $|k\rangle$ are around of corresponding label states, equally to the case that the red dots, green triangles and black boxes are around of blue star, black star and red star respectively, the test samples are successfully recognized. Obviously, the phenomenon appears in Fig. 4. It is seen clearly that most of data for the 4-layer case are more aggregated around corresponding label states than those for the 3-layer case. That is to say, red dots and green triangles in Fig. 4(C) are more aggregated around green star and black star respectively than the counterpart in Fig. 4(A). This phenomenon is in accordance with the above smaller accumulated MSE and higher recognition rate for the 4-layer case. Comparing them with the phenomenon in Fig. 2, the data of each type is not overlapped with others, which means that classifying is successfully completed. Although some data for different types are overlapped near the centers in Fig. 4(B) and (C), the phenomenon can be omitted because the centers locate at zero points.

## 4. Discussion and conclusion

In the training part of the above scheme, the distance between the unknown unit vector and the known unit vector has been obtained by performing the two measurements below Eq. (6). The computation of distance is exponential speedup and the reason is that the GHZ entanglement source has been used. We have made two vectors respectively connect with the bases $|0\rangle$ and $|1\rangle$ of ancillary qubit, as shown in Eq. (6). When the appropriate measurement basis is chosen, we can obtain the probability connecting with the MSE. When the dimensions of vectors are taken as N, we need $1+\log_2 N$ qubits, $2+2\log_2 N$ paths, $\log_2 N$ combiners and $2+\log_2 N$ detectors in our scheme as shown in Fig. 1(B). So, in order to get the MSE between two vectors with the dimension of N, we need quantum resources with $O(poly(\log N))$. Comparing them with the classical counterpart $O(poly(N))$ in classical computations [23], this is considered exponential speedup. Thus, the entanglement is the basis of speedup in our scheme and truly plays an important role in quantum information processing [63]. In fact, efficient computation of the distance using the GHZ entanglement state has been discussed in supervised and unsupervised machine learning algorithm, but it only focuses on two known vectors [23, 24]. The present results show that the quantum speed up scheme based on the GHZ entanglement state can also be realized for the case with unknown vectors.

In the part of QMFNN, we have realized quantized transformation in input layer, UU matrix, hidden layer and output layer corresponding to classical MFNN. However, due to the restrictions that is not strong or special for the connection between layers, weight matrix and UU matrix cannot be efficiently simplified like sparse matrix. For a matrix with the dimension of $d=2^n$, the required resource is $O(poly(d))$ and thus quantum speed up cannot be realized. This is similar with the quantum enhanced inference network [64] to some extent, which the precision-improved Hamiltonian simulation algorithm [65] is used by the linear combination of unitaries formalism [66]. It achieves that a term in the computational complexity reduces exponentially, albeit the overall complexity remains exponential in the number of nodes. However, if we use the quantum convolutional neural network instead of QMFNN in our scheme, the quantum speedup in computation can also be realized. This is because efficient computation by exponentially reducing the number of parameters can be implemented using the quantum convolutional neural network as described in Ref. [42].

In fact, some models of QDL have been discussed as described in Sec. I. Comparing them with the present scheme, we find there are some advantages for the present scheme. For example, the quantum Hopfield neural network only focuses on the special situations of information erasure [39]. However, our scheme is a general model that can be used in many situations. Besides, our scheme is based on feedforward neural networks, which is different from the feedback schemes such as quantum

Hopfield neural network [39] and quantum Boltzmann machine [34-36]. Most of feedforward neural networks are generally better than feedback neural networks in the abilities of classification and pattern recognition. Thus, the quantized feedforward scheme we proposed probably provides better results than the Hopfield and Boltzmann schemes for the two classes of task. Compared with another two schemes based on MFNN [43-44], our scheme encodes data into the probability amplitudes of quantum state, which is usually form in quantum computing and different from relative phase and Fock state. We think it is the reason that our scheme has the compatibility. Our training method for efficient computation of MSE can avoid the considerations about the knowledge of internal quantum state of the system [44] and thus provide a new way to finish optimizing quantum version of feedforward neural network. Compared with the quantum generalized scheme [45], our scheme has an efficient economy in the number of qubits and can relax the demands for experimental realization.

In conclusion, we have proposed a general scheme of QDL and realized full quantum process for the parts of QNN and training. Using GHZ entanglement state and the method of efficient computation of distance for an unknown unit vector and a known unit vector, exponential speed up in training has been realized. Based on the QMFNN, we have successfully recognized most of samples in Iris dataset. Besides, the scheme can be compatible with other models of QNN and improve the efficiency of computation. It is expected that the scheme will be beneficial to explore more efficient models of QDL in the future.

**Appendix**

In this appendix, we propose an experimental scheme of QDL with the dimension $N=4$. The setup is shown in Fig. 5(A). The GHZ entanglement resource is constructed by the degrees of freedom of polarization and OAM. Firstly, a pair of photons in polarized entangled state $|\psi\rangle_{A1} = \frac{1}{\sqrt{2}}(|H\rangle_A|H\rangle_1 + |V\rangle_A|V\rangle_1)$ is generated by the process of spontaneous parameter down conversion (SPDC) when the laser beam passes through the first beta-barium-borate crystal (BBO1). The subscripts A and 1 represents ancillary photon and the other photon labeled by 1. Photon 2 and 3 are generated by the second beta-barium-borate crystal (BBO2), which are used to perform the nonlinear transformation as described in Fig. 5(B). The entangled state $|\psi\rangle_{A1} = \frac{1}{\sqrt{2}}(|H\rangle_A|H\rangle_1|+2\rangle_1 + |V\rangle_A|V\rangle_1|-2\rangle_1)$ can be obtained when the photon 1 passes through the PBS

and two spiral phase plates (SPP). Here the two SPPs change the OAM of photon 1 from 0 to +2 and -2 for the horizontal and vertical polarization, respectively. When the horizontal polarization and the OAM of $l = +2$ are labeled as the computation basis 0, the vertical polarization and the OAM of $l = -2$ are labeled as 1, the above state corresponds to the 3-qubit GHZ state.

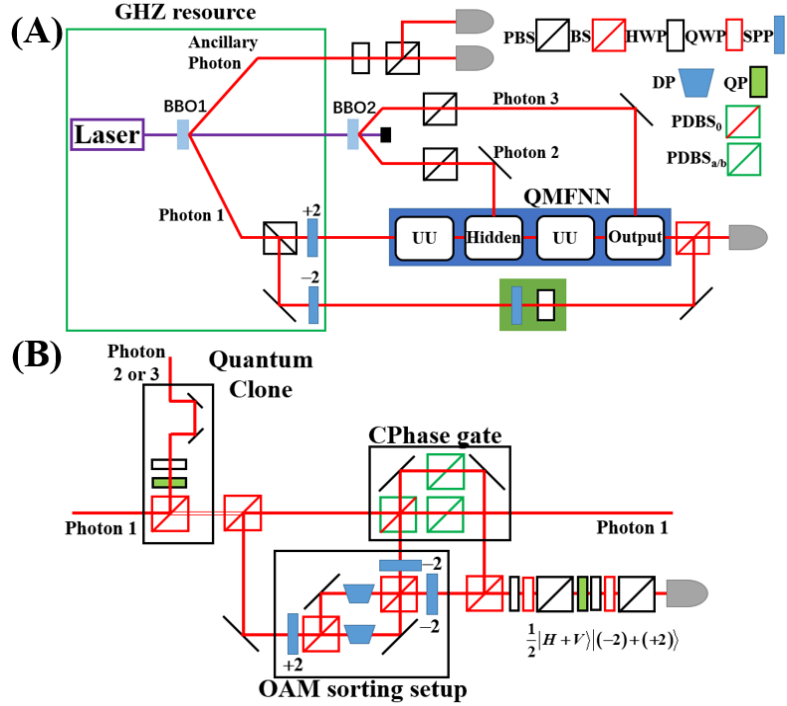

**Figure** 5. (A) Experimental scheme of QDL. The green area represents the operation of sample labels. The preparation of GHZ resource is shown in the green pane. The blue area represents QMFNN. UU, Hidden and Output correspond to universal unitary matrix, hidden layer and output layer, respectively. (B) The module of nonlinear transformation in the hidden layer. The module includes three parts: quantum clone, OAM sorting setup and control-phase (CPhase) gate. Legend of the main components: PBS – polarization beam splitter; BS – beam splitter; HWP – half-wave plate; QWP – quarter-wave plate; SPP – spiral phase plate; DP – dove prism; QP – Q-plate; $PDBS_0$ and $PDBS_{a/b}$ see Ref. [57]. ±2 means the variation of OAM when photon passes through the SPP.

In the part of QNN, the sub-state $|H\rangle_1|+2\rangle_1$ is transformed by QMFNN as shown in the blue area of Fig. 5(A). The preparation of input states can be viewed as a transformation of diagonal matrix and merged into the UU matrix to simplify experimental setup. Here, for the single photon with the degrees of freedom of polarization and OAM, the UU matrix can be realized by HWP, quarter-wave plate and Q-plate. The concrete scheme has been described in Ref. [53].

As for the nonlinear transformation when the sub-state is in the hidden layer, we use quantum clone combined with two-photon gate. The designed module is shown in Fig. 5(B). The module mainly includes three parts: quantum clone, OAM sorting setup and control-phase (CPhase) gate. In the quantum clone, the input photon with polarization and OAM can be cloned and two same photons are output. Photons 2 and 3 are used as ancillary photons in the process of clone. The clone setup has been discussed in Ref. [56]. Then we choose one of the two output photons to pass the OAM sorting setup. In this setup, the photon with $l=+2$ or $l=-2$ can be output in different paths by adjusting proper angles of dove prism (DP). The marks $\pm 2$ in Fig. 5(B) represent the variation of OAM when photon passes through the SPP. In the setup, we firstly change the OAM of photon from $l=+2$ and $l=-2$ to $l=+4$ and $l=0$ by using the SPP. After the interferometer, the OAM changes back to $l=+2$ and $l=-2$. The detailed discussion of OAM sorting setup has been given in Ref. [67]. Next, we choose the photon with $l=-2$ from the OAM sorting setup and the other cloned photon to pass the CPhase gate. Due to the orthogonalization for different OAMs, a $\pi$ phase shift occurs only for the situation that the two photons are vertical polarization and $l=-2$. The photon with $l=+2$ from the OAM sorting setup or with $l=-2$ from the CPhase gate is projected into the state $\frac{1}{2}|H+V\rangle|(+2)+(-2)\rangle$, and the other photon as the photon 1 proceeds following transformations. In the process, the nonlinear transformation is expressed as

$$\begin{pmatrix} \alpha_0 \\ \alpha_1 \\ \alpha_2 \\ \alpha_3 \end{pmatrix} \rightarrow \begin{pmatrix} \alpha_0(\alpha_0+\alpha_1+\alpha_2+\alpha_3) \\ \alpha_1(\alpha_0+\alpha_1+\alpha_2+\alpha_3) \\ \alpha_2(\alpha_0+\alpha_1+\alpha_2+\alpha_3) \\ \alpha_3(\alpha_0+\alpha_1+\alpha_2-\alpha_3) \end{pmatrix}. \quad (A1)$$

This means that the QMFNN can be realized by means of the above setup. In fact, there are other construction ways for such nonlinear transformation, like using the nonlinear sign-shift operation [54]. With the development of related researches, some new ways to realize the QDL may also be proposed.

In the part of training, the sub-state $|V\rangle_1|-2\rangle_1$ passes through the green area in Fig. 5(A). In such an area, the operation of sample labels can be implemented by a SPP and a HWP. For example, the state $|H\rangle_1|-2\rangle_1$ labeling one sample can be generated by rotating the HWP. To perform measurement, the certain basis for ancillary photon is realized by rotating HWP. The measurement for photon 1 is implemented by using the BS with the paths of the sample label and QMFNN. The MSE is obtained by coincidence counting with simple calculation. As for updating the independent parameters of UU matrix, we can change a tiny value for the parameters and repeat above process. According to Eq. (11), gradients are obtained by MSEs and used to update process.

In fact, the above scheme can be realized with current techniques. Every part of the scheme, like four-photon source [68], quantum clone [56], OAM sorting setup [67] and CPhase gate [57], has been demonstrated individually, which provides the basis of overall realization. When combining these parts, only two sequential two-photon gates are used, which is the state of the art in photon control [69]. Besides, three sequential two-photon gates can be realized equally by using hyperentanglement [70]. Although the quantum clone setup operates on two photons, one BS with two-photon interference is introduced which is simpler comparing with the two-photon gate [71]. As for the UU matrix, it can be realized by some optical elements in the same optical path.

## Acknowledgments

This work was supported by the National key R&D Program of China under Grant No. 2017YFA0303800 and the National Natural Science Foundation of China through Grants No. 11574031 and 61421001.